\documentclass[a4paper,11pt]{article}
\usepackage{pos}

\title{Phenomenology of a scotogenic neutrino mass model at 3-loops}

\author[a]{Asmaa~Abada}
\author[b]{Nicol\'{a}s Bernal}
\author[c,d,e]{Antonio E. C\'{a}rcamo Hern\'{a}ndez}
\author[d,e,f]{Sergey Kovalenko}
\author*[e,f]{T\'{e}ssio B. de Melo}
\author[g,h]{Takashi Toma}

\affiliation[a]{P\^ole Th\'eorie, Laboratoire de Physique des 2 Infinis Ir\`ene Joliot Curie (UMR 9012)\\
CNRS/IN2P3, 15 Rue Georges Clemenceau, 91400 Orsay, France\\}

\affiliation[b]{New York University Abu Dhabi, PO Box 129188, Saadiyat Island, Abu Dhabi, United Arab Emirates\\}

\affiliation[c]{Universidad T\'{e}cnica Federico Santa Mar\'{\i}a, Casilla 110-V, Valpara\'{\i}so, Chile\\}

\affiliation[d]{Centro Cient\'{\i}fico-Tecnol\'{o}gico de Valpara\'{\i}so, Casilla 110-V, Valpara\'{\i}so, Chile\\}

\affiliation[e]{Millennium Institute for Subatomic Physics at the High Energy Frontier (SAPHIR),\\
Fern\'andez Concha 700, Santiago, Chile\\}

\affiliation[f]{Center for Theoretical and Experimental Particle Physics - CTEPP,\\
Facultad de Ciencias Exactas, Universidad Andres Bello, Fernandez Concha 700, Santiago, Chile\\}

\affiliation[g]{Institute of Liberal Arts and Science, Kanazawa University, Kanazawa 920-1192, Japan\\}

\affiliation[h]{Institute for Theoretical Physics, Kanazawa University, Kanazawa 920-1192, Japan\\}

\emailAdd{asmaa.abada@ijclab.in2p3.fr}
\emailAdd{nicolas.bernal@nyu.edu}
\emailAdd{antonio.carcamo@usm.cl}
\emailAdd{sergey.kovalenko@unab.cl}
\emailAdd{tessiomelo@institutosaphir.cl}
\emailAdd{toma@staff.kanazawa-u.ac.jp}

\abstract{
\noindent
By extending the minimal scotogenic model with a spontaneously broken global symmetry $U(1)'$ and a preserved $\mathbb{Z}_2$ symmetry, we build a seesaw model for generating neutrino masses at three-loop level. The new particles have masses at the TeV scale and relatively large Yukawa couplings, which leads to sizable rates for charged lepton flavor violation processes, well within future experimental reach. The model is able to successfully explain the $W$ mass anomaly and provides a viable fermionic or scalar dark matter candidate, while satisfying all current constraints imposed by neutrinoless double-beta decay, charged-lepton flavor violation, and electroweak precision observables.
}

\FullConference{XVIII International Conference on Topics in Astroparticle and Underground Physics (TAUP2023)\\
 28.08-01.09.2023\\
University of Vienna\\}

\begin{document}
\maketitle

\section{Introduction}

Radiative seesaw models are examples of interesting and testable extensions of the SM to explain the light neutrino masses. 
In radiative seesaw models at 1-loop level, in order to successfully reproduce neutrino masses and mixing, one has to rely either on unnaturally small Yukawa couplings or on a very small mass splitting between the CP-even and CP-odd components of the neutral scalar mediators.
In this work, we propose an extended scotogenic model~\cite{Tao:1996vb,Ma:2006km} with moderate particle content, where light-active neutrino masses arise at the three-loop-level, providing a more natural explanation for the smallness of the neutrino masses. 
The model under consideration is consistent with the neutrino oscillation data and allows to successfully accommodate the measured dark matter (DM) relic abundance, as well as the constraints arising from charged-lepton flavor violation (cLFV), oblique parameters, in addition to being consistent with the observed $W$-mass anomaly.

In what follows, we introduce the model in Section~\ref{Sec:model}, providing a description of its field content and symmetries. We discuss some of its phenomenological aspects in Section~\ref{sec:cLFVobs}, in particular electroweak precision observables and cLFV processes, and summarize our findings in Section~\ref{Sec:conclusions}.

\section{Symmetries and particle content} \label{Sec:model}

\begin{figure}[!t]
    \begin{center}
        \includegraphics[width=0.45\textwidth]{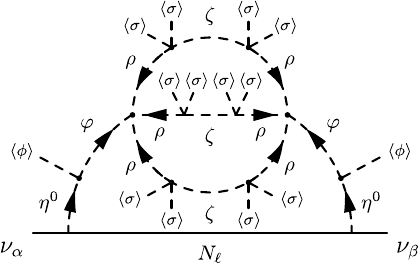} 
    \end{center}
    \caption{Scotogenic loop for light active neutrino masses where $\ell=1$, 2 and $\protect\alpha,\, \protect\beta = e,\, \protect\mu,\, \protect\tau$.}
    \label{Neutrinodiagram}
\end{figure}

We propose an extension of the SM with an augmented symmetry group, by the inclusion of a spontaneously broken global symmetry $U(1)'$ and a preserved discrete symmetry $\mathbb{Z}_2$.
The SM particle content is extended by an inert scalar doublet $\eta$ and two RH neutrinos $N_{R_k}$, which have a non-trivial $\mathbb{Z}_2$ charge, as in the scotogenic model. The lightest state among the electrically neutral components of $\eta$ and the two states $N_{R_{k}}$ is thus a viable DM candidate.
In addition, four electrically neutral scalar singlets $\sigma$, $\rho$, $\varphi$, $\zeta$ are also included, all of them odd under $\mathbb{Z}_2$, except for $\sigma$, which is responsible for breaking the $U(1)'$ symmetry at the TeV scale. 
Given these symmetries and particle spectrum, the allowed charged-fermion and neutrino Yukawa interactions are, 
\begin{align}
    -\mathcal{L}_Y \supset y_{u \phi}^{ij}\, \bar{q}_{iL} \widetilde{\phi} u_{jR} + y_{d \phi}^{ij}\, \bar{q}_{iL} \phi d_{jR} + y_{l \phi}^{ij}\, \bar{\ell}_{iL} \phi \ell_{R_j} + y_\eta^{ik}\, \bar{\ell}_{iL} \widetilde{\eta} N_{R_{k}} + M_{N_{R}}^{kr}\, \bar{N}_{R_{k}} N_{R_{r}}^C + \mathrm{H.c.}
\label{eq:lagrangian2}\ ,\end{align}
while in the scalar sector, the scalar potential include the terms,
\begin{align}
    V& \supset \lambda _{15}\left( \rho \zeta \sigma ^{2}+\mathrm{H.c.}\right) + \lambda_{14}\left( \varphi \rho ^{3}+\mathrm{H.c.}\right) + A\left[ (\eta ^{\dagger }\phi )\varphi +\mathrm{H.c.}\right] .
    \label{eq:sacalar-potential}  
\end{align}
Together, these interactions are responsible for the 3-loop diagram shown in Fig. \ref{Neutrinodiagram}, which is the leading contribution to the neutrino masses, since the tree-, 1-loop- and 2-loop-level contributions are forbidden by the symmetries of the model.

\section{Phenomenological implications} \label{sec:cLFVobs}

\begin{figure}[b]
    \centering
    \includegraphics[width=0.48\linewidth]{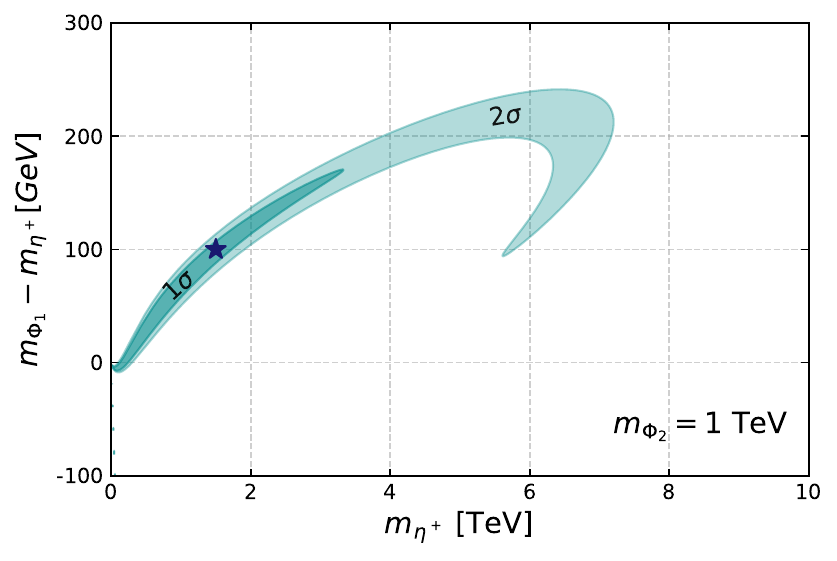}\quad
    \includegraphics[width=0.48\linewidth]{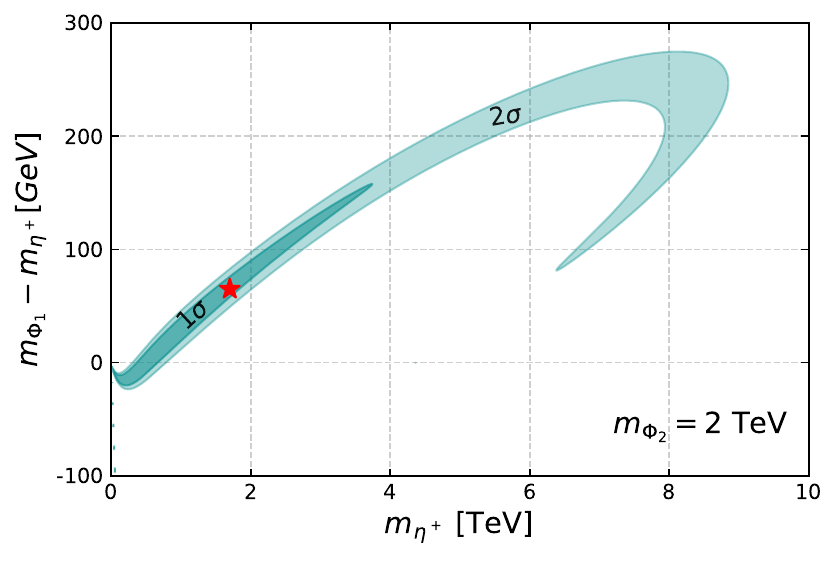}
    \caption{The $1$$\protect\sigma$ and $2$$\protect\sigma$ regions which accommodates the CDF measurement of the $W$ mass.
    }
    \label{stuCDFplot}
\end{figure}

\paragraph{Oblique parameters and $W$ boson mass}

The presence of extra scalars in our model affect the oblique corrections of the SM, which are parameterized in terms of the well-known quantities $T$, $S$ and $U$. Since their values are measured in high-precision experiments, they act as constraints on our model. Furthermore, the measurement of the $W$ gauge boson mass by the CDF collaboration~\cite{CDF:2022hxs}, can be interpreted as an indication of non-trivial $S$, $T$, and $U$ values, according to: 
\begin{equation}
    M_W^2 = \left(M_W^2\right)_\text{SM} + \frac{\alpha_\text{EM}\left(M_Z\right) \cos^2\theta_W\, M_Z^2}{\cos^2\theta_W - \sin^2\theta_W} \left[-\frac{S}{2} + \cos^2\theta_W\, T + \frac{\cos^2\theta_W - \sin^2\theta_W}{4\, \sin^2\theta_W}\, U\right] .
\end{equation}
In Fig.~\ref{stuCDFplot}, we plot the $1$-$\protect\sigma$ and $2$-$\protect\sigma$ parameter space regions in which the CDF result can be explained. 
We note that the CDF measurement can accommodate scalar masses in the TeV scale, provided that the mass splitting among the charged and neutral scalars is not larger than a few hundred GeV. In this scenario, in addition to explaining the anomaly and avoiding the current experimental constraints, the model becomes highly predictive, providing signatures in cLFV experiments, as detailed below.

\begin{figure}[t!]
    \centering
    \includegraphics[width=0.62\linewidth]{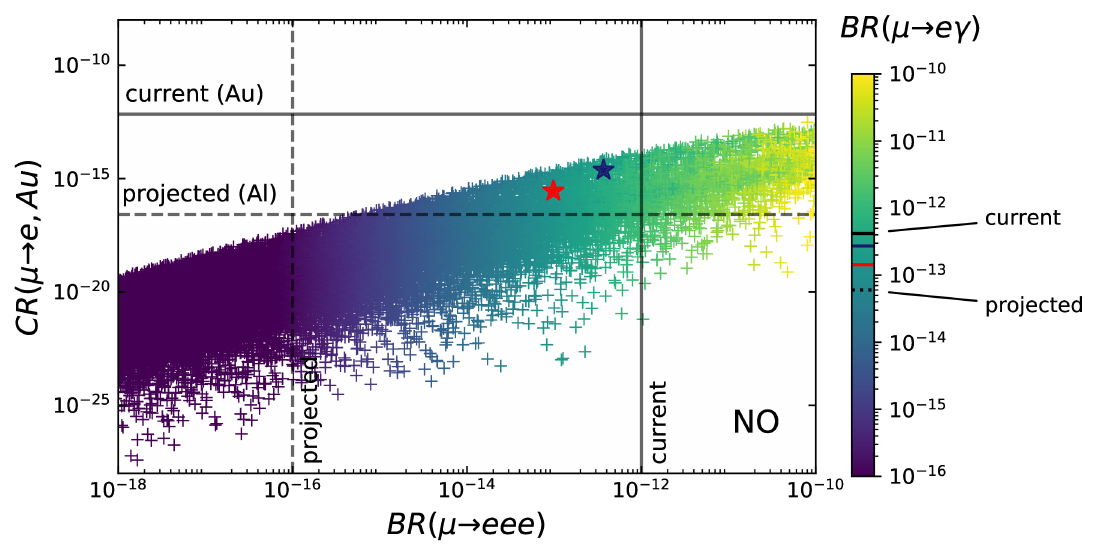}
    \caption{Correlation between the cLFV processes $\mu \to e \gamma$, $\mu \to eee$ and $\mu \to e$ conversion in gold nuclei. The current (projected) upper bounds are indicated by the black full (dashed) lines.}
    \label{LFVplot3}
\end{figure}

\paragraph{Charged lepton flavor violating observables}

The radiative cLFV decays $\mu \rightarrow e\gamma $ and $\mu \to e e e $, and $\mu-e$ conversion process in atomic nuclei, are excellent probes to test our model. They arise at 1-loop level from the exchange of charged scalars $\eta ^\pm$ and RH neutrinos $N_{R_k}$.
We perform a random scan over the parameters of the model and calculate the corresponding cLFV rates. 
We use an adapted Casas-Ibarra parameterization, so that all the calculated points reproduce the observed neutrino masses and mixing.
The result is shown in Fig.~\ref{LFVplot3}, assuming normal ordering (similar results are obtained for inverted ordering). 
It is apparent from Fig.~\ref{LFVplot3} that the high precision expected for future $\mu-e$ conversion and $\mu \to e e e$ experiments will allow to probe a large portion of the parameter space of the model. In Fig.~\ref{LFVplot3} we also show two benchmark points, marked as the blue and red stars (which have counterparts in Fig.~\ref{stuCDFplot}). In these points, all the current constraints are satisfied, the $W$ mass anomaly is explained and observable rates in cLFV future experiments are predicted. They are concrete examples of the potential and predictivity of the model.

\section{Conclusion} \label{Sec:conclusions}

We have constructed a 3-loop radiative seesaw model that produces the tiny active neutrino masses and accommodates a fermionic or scalar DM candidate. The 3-loop suppression allows the new particles to have masses in the TeV scale without fine-tuning the Yukawa couplings. We have shown that the model is capable of explaining the $W$ mass anomaly and leads to interesting phenomenology, especially on cLFV processes, which are within the sensitivity of future facilities.

\section*{Acknowledgments}

\noindent
The results here presented are based on Ref.~\cite{Abada:2022dvm}. 
TBM thank organizers of TAUP2023 conference for the opportunity to present this work.
NB received funding from the Spanish FEDER/MCIU-AEI under grant
FPA2017-84543-P. AECH and SK are supported by ANID-Chile FONDECYT 1210378, 1230160, ANID PIA/APOYO AFB220004, and ANID Programa Milenio code ICN2019$\_$044. TBM acknowledges CNPq (grant No. 164968/2020-2) and ANID-Chile FONDECYT (grant No. 3220454) for financial support.
This project has received funding and support from the European Union's Horizon 2020 research and innovation programme under the Marie Sk{\l}odowska-Curie grant agreement No.~860881 (H2020-MSCA-ITN-2019 HIDDeN). 
This work was supported by the JSPS Grant-in-Aid for Scientific Research KAKENHI Grant No. JP20K22349 (TT).

\bibliographystyle{unsrt}
\bibliography{biblio}

\end{document}